\documentclass[aps,pre,twocolumn,amsmath,amssymb,floatfix]{revtex4}
\usepackage[T1]{fontenc}
\usepackage[latin1]{inputenc}
\usepackage{graphicx}
\usepackage{float}

\providecommand*\femlab{\textsc{Femlab}}
\providecommand*\matlab{\textsc{Matlab}}
\providecommand*\reynold{\ensuremath{\mathit{Re}}}

\providecommand*\capillary{\ensuremath{\mathit{Ca}}}

\providecommand*\br{\ensuremath{\mathbf{r}}}
\providecommand*\bu{\ensuremath{\mathbf{u}}}
\providecommand*\ba{\ensuremath{\mathbf{a}}}
\providecommand*\bF{\ensuremath{\mathbf{F}}}
\providecommand*\bV{\ensuremath{\mathbf{V}}}
\providecommand*\bn{\ensuremath{\mathbf{n}}}
\providecommand*\da{\ensuremath{d_a}}
\providecommand*\bU{\ensuremath{\mathbf{U}}}

\providecommand*\bsigma{\ensuremath{\boldsymbol{\sigma}}}
\providecommand*\bGamma{\ensuremath{\boldsymbol{\Gamma}}}
\providecommand*\bnabla{\ensuremath{\boldsymbol{\nabla}}}
\providecommand*\order[1]{\mathcal{O}(#1)}

\providecommand*\determinant[1]{\left|#1\right|}

\providecommand*\stau[1][]{\Big(x(s,\tau#1),y(s,\tau#1)\Big)}

\makeatletter
\def\DIfF^#1{\mathop{\mathrm{\mathstrut d}}%
 \nolimits^{#1}\gobblespace}
\providecommand*{\diff}{\@ifnextchar^{\DIfF}{\DIfF^{}}}
\def\gobblespace{\futurelet\diffarg\opspace}
\def\opspace{%
 \let\DiffSpace\!%
 \ifx\diffarg(%
  \let\DiffSpace\relax
 \else
 \ifx\diffarg[%
  \let\DiffSpace\relax
 \else
 \ifx\diffarg\{%
  \let\DiffSpace\relax
 \fi\fi\fi\DiffSpace}
\makeatother

\providecommand*\deriv[3][]{\frac{\diff^{#1}#2}{\diff #3^{#1}}}
\providecommand*\pderiv[3][]{\frac{\partial^{#1}#2}{\partial #3^{#1}}}

\DeclareMathOperator\sign{sign}
\graphicspath{{./figures/}}
\begin{document}
\title{Particle motion in microfluidics simulated using a\\
\femlab\ implementation of the level set method}
\author{Martin Heller}
\author{Henrik Bruus}
\affiliation{%
 MIC -- Department of Micro and Nanotechnology, DTU bldg.~345 east\\
 Technical University of Denmark, DK-2800 Kongens Lyngby, Denmark
}
\date{\today}

\begin{abstract}
We implement the level set method for numerical simulation of
the motion of a suspended particle convected by the fluid flow
in a microchannel. The method automatically cope with the
interactions between the particle and the channel walls. We
apply the method in a study of particles moving in a channel
with obstacles of different shapes. The generality of the method
also makes it applicable for simulations of motion of particles
under influence of external forces.
\end{abstract}

\maketitle

\section{Introduction}
\label{sec:introduction}%
In recent years numeral lab-on-a-chip systems have been developed
to analyze biological samples. Many  of these systems rely on
handling of particles and cells comparable in size to the
dimensions of the channels containing them. Examples of such
microsystems are bumperarrays or
DEP-systems~\cite{bib:chou,bib:duke,bib:huang02,bib:huang04}

It is a major challenge in theoretical microfluidics to study
the dynamics of particles of finite size when they are convected
by a fluid flow. Especially problematic is the forces appearing
during collisions of the particles with the walls of the
channel.

The level set method \cite{bib:sethian} is well suited to cope
with these problems. By introducing a hypersurface
$\phi(\br,t)$, the particle interface is represented as the zero
level set $\phi(\br,t)=0$. The major advantage of the method is
that this zero level set can be calculated implicitly instead of
explicit tracking of the points on the interface.

The manuscript is organized as follows: In Sec.~\ref{sec:goveqns}
we state the equations governing the dynamics of the system and in
Sec.~\ref{sec:levelset} we derive the level set formulation for
the tracked interface. The implementation of the method in the
numerical simulation tool \femlab\ is described in
Sec.~\ref{sec:femlab} and we present results of a test study in
Sec.~\ref{sec:results}. Finally, we evaluate the method in
Sec.~\ref{sec:conclusion} and give suggestions to future areas of
usage.

\section{Governing equations}
\label{sec:goveqns} We consider microfluidic systems. Hence the
characteristic length scales of channels are of the order of
10~$\mu$m which is well beyond the intermolecular distances
characteristic of the fluids involved. Thus the continuum
hypothesis applies. Moreover, in these systems the flow velocities
are much smaller than the propagation of pressure (the speed of
sound). We can therefore consider the fluids to be incompressible
and the continuity condition
\begin{equation}\label{eqn:continuity}
 \bnabla\cdot\bu = 0
\end{equation}
holds true for the velocity field \bu\ of the fluid.

Consider a domain $\Omega$ consisting of two subdomains $\Omega_1$
and $\Omega_2$ with surfaces $\partial\Omega_1$ and
$\partial\Omega_2$, respectively. The common boundary between
$\Omega_1$ and $\Omega_2$ is the interface $\Gamma$ which we want
to evolve.

The rate of change of the momentum of the fluid is given by
$\int_\Omega\!\rho\frac{\mathrm{D}\bu}{\mathrm{D}t}\diff\br$
involving the substantial time derivative of \bu. The change in
momentum arises from the forces acting on the volume of fluid. In
a microfluidic system we can neglect gravity and the only force
$\bF_\sigma$ acting on a volume of fluid $\Omega$ stems from the
stresses $\bsigma$ exerted by the surrounding liquid on the
surface $\partial\Omega$,
\begin{equation}
\bF_\sigma=\int_{\partial\Omega}\!\bsigma\cdot\diff\ba\text{,}
\end{equation}
where $\bsigma$ is the stress tensor modelled by
\begin{align}\label{eq:stress}
\sigma_{ij}=-p\delta_{ij}+\eta\left(\partial_j u_i+\partial_i
u_j\right)\text{.}
\end{align}
Newton's second law therefore takes the form
\begin{equation}
\int_\Omega\!\rho\frac{\mathrm{D}\bu}{\mathrm{D}t}\diff\br
=\int_{\partial\Omega}\!\bsigma\cdot\diff\ba\text{.}
\end{equation}
The right hand side of this equation can be split up in three
integrals; two parts for each of the boundaries of the two
subdomains and one along the common interface
\begin{align}\label{eq:newtons second law for two viscid system}
\int_\Omega\!\rho\frac{\mathrm{D}\bu}{\mathrm{D}t}\diff\br
&=\int_{\partial\Omega_1}\!\bsigma\cdot\diff\ba
+\int_{\partial\Omega_2}\!\bsigma\cdot\diff\ba
+\int_{\Gamma}\![\bsigma\cdot\diff\ba]\nonumber\\
&=\int_{\Omega_1}\!\bnabla\cdot\bsigma\diff\br
+\int_{\Omega_2}\!\bnabla\cdot\bsigma\diff\br
+\int_{\Gamma}\!\gamma\kappa\diff\ba\text{,}
\end{align}
In the second equality we have used Gauss' theorem as well as
the Young--Laplace law relating the pressure drop
$[\bsigma\cdot\diff\ba]$ across the interface $\Gamma$ to the
surface tension $\gamma$ and average curvature $\kappa$.

To facilitate numerical computation it is desirable to rewrite the
last integral in Eq.~\eqref{eq:newtons second law for two viscid
system} as a volume integral like the rest of the terms. This can
be achieved by introducing a level set function $\phi(\br,t)$ as
we will show in the following.

\section{The level set method}
\label{sec:levelset} Following Ref.~\cite{bib:chang} we introduce
a level set function $\phi(\br,t)$ with the properties
\begin{equation}\label{eq:level set definition}
\begin{cases}
\phi(\br,t)>0\text{,} & \br\in\Omega_1\text{,}\\
\phi(\br,t)=0\text{,} & \br\in\Gamma\text{,}\\
\phi(\br,t)<0\text{,} & \br\in\Omega_2\text{.}
\end{cases}
\end{equation}
This function uniquely defines the interface as $\Gamma(t) =
\{\br|\phi(\br,t)=0\}$ and permits us to distinguish each
subdomain by the sign of $\phi$. We also introduce a transverse
level set function $\psi(\br,t)$ such that
\begin{equation}
\bnabla\phi\cdot\bnabla\psi=0,\quad\vert\bnabla\psi\vert\neq0\text{.}
\end{equation}
We show in Appendix~\ref{app:proofs} that it is possible to
construct such level set functions. In the following we consider
a two dimensional system, but the method is applicable in higher
dimensions also. We can construct a global
orientation-preserving diffeomorphism that maps $\Omega \mapsto
\Omega'$ through the variable change
\begin{subequations}\label{eq:variable change}
\begin{align}
x'&=\psi(x,y) \\
y'&=\phi(x,y)\text{.}
\end{align}
\end{subequations}
We denote partial derivatives with indices,
e.g.,~$\psi_x\equiv\partial_x\psi$. The change of variables
Eqs.~\eqref{eq:variable change} is area preserving because the
Jacobian is non-zero,
\begin{align}\label{eq:change of variables}
\left|\pderiv{(\psi,\phi)}{(x,y)}\right|
=(\phi_y,-\phi_x)\cdot(\psi_x,\psi_y)
=|\bnabla\phi||\bnabla\psi|\neq 0\text{,}
\end{align}
where we assume that $\psi$ is constructed such that
$\bnabla\psi$ is parallel to the tangent direction and therefore
$-\hat{\bnabla}\phi||\bnabla\psi$.

Furthermore we introduce a parameterization
$\big(\overline{x}(s),\overline{y}(s)\big)$ of $\Gamma$, where $s$ is an
arc-length variable. Using this parameterization an infinitesimal
change in $x'$ along $\Gamma$ is given by
\begin{equation}
\diff x'|_{\phi=0}=\vert\bnabla\psi\vert\diff s\text{,}
\end{equation}
where we have utilized the above assumption that the gradient of
$\psi$ is parallel to the tangent direction. With the above
definitions we can rewrite the surface integral in
Eq.~\eqref{eq:newtons second law for two viscid system} as
\begin{align}\label{eq:pressure drop over gamma transformed}
\begin{split}
\int_\Gamma\!\gamma\kappa\diff\ba
&=\int_{\phi=0}\!\gamma\kappa\bn\diff s\\
&=\int_{\phi=0}\!\gamma\kappa\frac{\bnabla\phi}{\vert\bnabla\phi\vert}\frac{1}{\vert\bnabla\psi\vert}\diff x'\\
&=\int_{\Omega'}\!\gamma\kappa\delta(y')\frac{\bnabla\phi}{\vert\bnabla\phi\vert}\frac{1}{\vert\bnabla\psi\vert}\diff x'\diff y'\text{,}
\end{split}
\end{align}
where we have used that the normal \bn\ to the interface can be
written as $\bnabla\phi/|\bnabla\phi|$. Using Eq.~\eqref{eq:change
of variables} for changing variables, Eq.~\eqref{eq:pressure drop
over gamma transformed} becomes
\begin{equation}\label{eq:pressure drop over gamma rewritten}
\int_\Gamma\!\gamma\kappa\diff\ba
=\int_\Omega\!\gamma\kappa\delta(\phi)\bnabla\phi\diff x \diff
y\text{.}
\end{equation}
Inserting Eq.~\eqref{eq:pressure drop over gamma rewritten} into
Eq.~\eqref{eq:newtons second law for two viscid system} yields
\begin{equation}
\int_\Omega\!\rho\frac{\mathrm{D}\bu}{\mathrm{D}t}\diff\br
=\int_{\Omega}\!\left[\bnabla\cdot\bsigma+\gamma\kappa\delta(\phi)\bnabla\phi\right]\diff\br\text{.}
\end{equation}
This must hold true for any volume $\Omega$. Hence
\begin{align}\label{eq:the level set equation}
\rho\left[\partial_t\bu+(\bu\cdot\bnabla)\bu\right]
&=\bnabla\cdot\bsigma+\gamma\kappa\delta(\phi)\bnabla\phi\text{,}
\end{align}
which is the level set formulation of the Navier--Stokes equation.

In order to have the system completely described by dynamical
equations we finally need an equation describing the evolution of
the zero level set. We only need to consider the movement of the
zero level set because this is the only part of the level set
function with a physical interpretation. Evolving the equation
$\phi(\br,t)=0$ in time defines the movement of the front.
Differentiating with respect to time yields
$\deriv{}{t}\phi(\br,t)=0$ which is written as
\begin{equation}
\partial_t\phi(\br,t)+\bV\cdot\bnabla\phi(\br,t)=0\text{,}
\end{equation}
where $\bV=\left.\deriv{\br}{t}\right|_{\br\in\Gamma}$ is the
velocity of the zero level set.

Requiring the velocity field to be continuous leads to
$\bV=\bu$, and the evolution equation for $\phi$ becomes
\begin{align}\label{eq:evolution equation for the level set function}
\phi_t+\bu\cdot\bnabla\phi=0\text{.}
\end{align}

\section{\label{sec:femlab}\femlab\ implementation}
One of the great advantages of the level set formulation is that
it does not track the interface explicitly but rather capture it
implicitly. Thereby we avoid to introduce explicit forces from
the walls during collisions as they enter implicitly through the
stress tensor $\bsigma$ and the no-slip boundary condition on
the velocity field \bu. Furthermore, several numerical tools are
available for solving the dynamical system. In this section we
describe how we have implemented the level set method in the
finite element software package \femlab~\cite{bib:comsol}. We
have used the \femlab\ scripting language trough a \matlab\
interface in the general PDE mode. Here the PDEs are given by
\begin{subequations}
\begin{align}
\da\deriv{\bU}{t}&+\bnabla\cdot\bGamma=\bF&\text{in $\Omega$}
\intertext{in terms of the variable vector $\bU$, the current
tensor $\bGamma$ and the generalized source terfield $\bF$. The
boundary conditions take the form}
-n_j\Gamma_{lj} &= G_l + \pderiv{R_m}{U_l}\mu_m&\text{on $\partial\Omega$}\\
0&=R_m&\text{on $\partial\Omega$,}
\end{align}
\end{subequations}
where the index $l$ is the variable counter, $m$ is the
constraint number (the number of boundaries) and $j$ is the
number space dimension number. The Lagrange multipliers $\mu_m$
are chosen by \femlab\ in order to fulfill the constraints,
while the scalars $F_l$, $G_l$ and $R_m$ are given by the
physics of the problem.

\subsection{\label{subsec:ns-femlab}Navier--Stokes equation in \femlab}
Introducing the characteristic length scale $L_0$, velocity scale
$U_0$, density $\rho_0$, viscosity $\eta_0$ and surface tension
$\gamma_0$ we can express the physical quantities as a
dimensionless number times the characteristic scale. Denoting the
nondimensional quantities by a tilde we simply have
\begin{equation}
\begin{aligned}
\mathbf{r}&=L_0\tilde{\mathbf{r}}\text{,} &
\bu&=U_0\tilde{\bu}\text{,} &
\rho&=\rho_0 \tilde{\rho}\text{,}\\
\eta&=\eta_0 \tilde{\eta}\text{,} &
\gamma&=\gamma_0\tilde{\gamma}\text{.}\label{eq:dimless1}
\end{aligned}
\end{equation}
Similarly we can define the characteristic pressure and timescale
as relations between the chosen characteristic parameters
\begin{align}\label{eq:dimless2}
p=\frac{\eta_0 U_0}{L_0}\tilde{p}\text{,}\qquad
t=\frac{L_0}{U_0}\tilde{t}\text{.}
\end{align}
Substituting Eqs.~\eqref{eq:dimless1} and~\eqref{eq:dimless2}
into the Navier--Stokes equation~\eqref{eq:the level set
equation} yields
\begin{align}\label{eq:navier stokes equation on dimensionless form}
\reynold\tilde{\rho}\left[\partial_{\tilde{t}}\tilde{\bu}
+(\tilde{\bu}\cdot\tilde{\bnabla})\tilde{\bu}\right]
=\tilde{\bnabla}\cdot\tilde{\bsigma}
+\frac{1}{\capillary}\tilde{\gamma}\tilde{\kappa}\delta(\phi)\tilde{\bnabla}\phi\text{.}
\end{align}
Here the Reynolds number $\reynold=\rho_0 U_0 L_0/\eta_0$ is the
ratio between inertial forces and viscous forces and the
Capillary number $\capillary=\eta_0 U_0/\gamma_0$ is the ratio
between viscous forces and the surface tension forces.

Rearranging the terms in Eq.~\eqref{eq:navier stokes equation on
dimensionless form} we find
\begin{align}
\reynold\tilde{\rho}\partial_{\tilde{t}}\tilde{\bu}
-\tilde{\bnabla}\cdot\tilde{\bsigma}
=\frac{1}{\capillary}\tilde{\gamma}\tilde{\kappa}\delta(\phi)\tilde{\bnabla}\phi
-\reynold\tilde{\rho}(\tilde{\bu}\cdot\tilde{\bnabla})\tilde{\bu}\text{,}
\end{align}
which is seen to be on the \femlab\ general form if
\begin{subequations}
\begin{align}
\da&=\reynold\tilde{\rho}\text{,}\\
\bGamma&=-\tilde{\bsigma}\text{,}\\
\bF&=-\reynold\tilde{\rho}(\tilde{\bu}\cdot\tilde{\bnabla})\tilde{\bu}
+\frac{1}{\capillary}\tilde{\gamma}\tilde{\kappa}\delta(\phi)\tilde{\bnabla}\phi\text{,}\\
\bU_\bu&=\tilde{\bu}\text{.}
\end{align}
\end{subequations}
The density $\tilde{\rho}$, viscosity $\tilde{\eta}$ and the
curvature of the front $\tilde{\kappa}$ are defined as auxiliary
functions of the level set function $\phi$. In a system with two
immiscible incompressible fluids (or a particle in a fluid) the
density and viscosity are constant on each side of the
interface. We can therefore define the dimensionless density and
viscosity as
\begin{align}
\tilde{\rho}&=1+H(\phi)\left(\frac{\rho_1}{\rho_2}-1\right)\\
\intertext{and}
\tilde{\eta}&=1+H(\phi)\left(\frac{\eta_1}{\eta_2}-1\right)\text{,}
\end{align}
where $H(\phi)$ is a Heaviside function defined as
\begin{align}
H(\phi) =
\begin{cases}
1\text{,} & \phi\in\Omega_1\text{,} \\
0\text{,} & \phi\in\Omega_2\text{.}
\end{cases}
\end{align}
Setting $\rho_0 = \rho_2$ ensures that the density of the fluid
is $\rho_1$ and $\rho_2$ in $\Omega_1$ and $\Omega_2$,
respectively. Similarly setting $\eta_0 = \eta_2$ makes the
viscosity of the fluid $\eta_1$ and $\eta_2$ in $\Omega_1$ and
$\Omega_2$, respectively.

The curvature of the zero level set is given by
\begin{align}
\kappa(\phi)=\bnabla\cdot\bn=
\bnabla\cdot\left(\frac{\bnabla\phi}{\vert\bnabla\phi\vert}\right)\text{,}
\end{align}
where $\bn=\bnabla\phi/\vert\bnabla\phi\vert$ is a unit normal
vector to the interface~\cite{bib:osher,bib:sethian}.

When solving the system numerically the abrupt change in density
and viscosity across the interface causes numerical
instabilities to occur. In order to avoid this we substitute
$H(\phi)$, $\delta(\phi)$ and $\sign(\phi)$ with the smeared out
versions $H_\epsilon(\phi)$, $\delta_\epsilon(\phi)$ and
$\sign_\epsilon(\phi)$ defined as
\begin{subequations}
\begin{align}
H_\epsilon(\phi)&=\frac{1}{2}
+\frac{1}{2}\tanh\left(\frac{\phi}{\epsilon}\right)\text{,}\label{eq:smhs}\\
\delta_\epsilon(\phi)&=H_\epsilon'(\phi)=\frac{1}{2\epsilon}
-\frac{1}{2\epsilon}\tanh^2\left(\frac{\phi}{\epsilon}\right)\text{,} \label{eq:smdelta}\\
\sign_\epsilon(\phi)&=\tanh\left(\frac{\phi}{\epsilon}\right)\text{.}\label{eq:smsign}
\end{align}
\end{subequations}
This implies that the interface has a finite thickness
$\Gamma_\epsilon$ approximately given by
\begin{align}
\Gamma_\epsilon=\frac{2\epsilon}{|\bnabla\phi|}\text{.}
\end{align}

\subsection{\label{subsec:cont-femlab}The continuity equation in \femlab}
The dimensionless form of the continuity equation is
\begin{align}
0=\tilde{\bnabla}\cdot\tilde{\bu}\text{,}\label{eq:continuity
equation on dimensionless form}
\end{align}
which is entered into \femlab\ by choosing
$\bF=\tilde{\bnabla}\cdot\tilde{\bu}$, $\bGamma=\boldsymbol{0}$,
$\da=0$ and $U_p=\tilde{p}$.

\subsection{\label{subsec:ls-femlab}The level set equation in \femlab}
The nondimensionalized form of the convection equation for the
zero level set is
\begin{align}
\phi_{\tilde{t}}+\tilde{\bu}\cdot\tilde{\bnabla}\phi=0\text{,}\label{eq:level
set equation on dimensionless form strong form}
\end{align}
which can be rearranged to
\begin{align}
\phi_{\tilde{t}}=-\tilde{\bu}\cdot\tilde{\bnabla}\phi\label{eq:level
set equation on dimensionless form}
\end{align}
and implemented in \femlab\ by setting
$\bF=-\tilde{\bu}\cdot\tilde{\bnabla}\phi$,
$\bGamma=\boldsymbol{0}$, $\da=1$ and $U_\phi=\tilde{\phi}$.

\begin{table}[H]
\centering \caption{The parameter values used in the simulation
of the test case.} \label{tab:parametervalues}
\begin{tabular}{ll@{$\,=\;$}r@{$\,\times\,$}l}
\hline
Reynolds number & \reynold & $1$ & $10^{-3}$ \\
Capillary number & \capillary & $1$ & $10^6$ \\
Density & $\rho_0$ & $1$ & $10^3$~kg$\:$m$^{-3}$ \\
Viscosity & $\eta_0$ & $1$ & $10^{-1}$~Pa$\:$s \\
Obstacle size & $l$ & $6$ & $10^{-6}$~m \\
Particle radius & $r_p$ & 3 & $10^{-6}$~m \\
Pressure drop & $\Delta p$ & $1.2$ & $10^{-3}$~Pa \\
Time step & $\Delta t$ & $5$ & $10^{-2}$~s \\
Mesh element size & \texttt{hmesh} & $1.1$ & $10^{-6}$~m \\
Thickness parameter & $\epsilon$ & $0.5$ & \texttt{hmesh} \\
\hline
\end{tabular}
\end{table}
\subsection{\label{subsec:reinitialization-femlab}Reinitialization of the level set function}
It is necessary to maintain a uniform thickness of the interface
throughout the calculations. This requires that the gradient of
the level set function is constant within a region around the
interface $|\phi|<\epsilon$. This is not automatically fulfilled.
The time evolution of any level set $\phi(\br,t) = C$ is given by
the level set Eq.~\eqref{eq:evolution equation for the level set
function}. This means that the height of the level set function
will remain constant, but it does not ensure that the gradient
does not change. Thus in order to keep a fixed interface thickness
we need to reinitialize the level set function without changing
the zero level set.

In principle we can use any function that fulfills
Eq.~\eqref{eq:level set definition}, since only the zero level set
has a physical interpretation. But requiring the interface
thickness to be fixed constrains the gradient of $\phi$ to be
fixed in a region around the interface. A choice of $\phi(\br,t)$
that fulfills these requirements is the signed distance function,
where the distance is the shortest distance $d(\br)$ from a point
to the interface
\begin{align}
d(\br)=\pm\min(\vert\br-\br_\Gamma\vert),
\end{align}
$\br_\Gamma$ being the points on the interface. The plus sign
applies if $\br\in\Omega_1$ and the minus sign if
$\br\in\Omega_2$. The length of the gradient for this particular
choice of level set function is
\begin{align}
|\bnabla\phi|=1\text{.}
\end{align}

We have implemented two different reinitilization procedures. One
simple reinitialization procedure where we recalculate the
level set function at every time step and one using the
reinitialization equation suggested by Sussmann, Smereka and Osher
\cite{bib:sussmann}
\begin{align}
\partial_\tau\psi(\mathbf{r},\tau)
=\sign(\phi)\big(1-\vert\bnabla\psi(\mathbf{r},\tau)\vert\big)\text{,}
\end{align}
with the initial condition $\psi(\mathbf{r},0)=\phi$ and $\tau$
being a pseudotime. The steady state solution to this equation
is the reinitialized level set function. Because numerical
oscillations can occur if the sign of $\phi$ changes abruptly at
the interface it is necessary to use the smeared out sign function
given in Eq.~\eqref{eq:smsign}.

The reinitialization equation is already on a form suitable for
implementation in \femlab. Simply letting $F$ equal the right
hand side of the equation and setting $\da=1$ and
$\bGamma=\boldsymbol{0}$ with $U_\psi=\psi$ does the trick.

To avoid mass loss during the reinitialisation procedure we have
put a constraint on the solution: the volume of the particle
must be constant at all time. This is done in \femlab\ via the
field \texttt{fem.equ.constr} where we constrain the difference
between the integrals of the smeared out Heaviside function
$H_\epsilon(\psi)$ at time $\tau$ and the smeared out Heaviside
function $H_\epsilon(\phi)$ at time $t=0$ to be zero. The
integrals are computed by using the integration coupling
variables in \femlab.

\section{\label{sec:model system}Model system and setup}
\begin{figure}[t]
\centering
\includegraphics*[width=\columnwidth,clip]{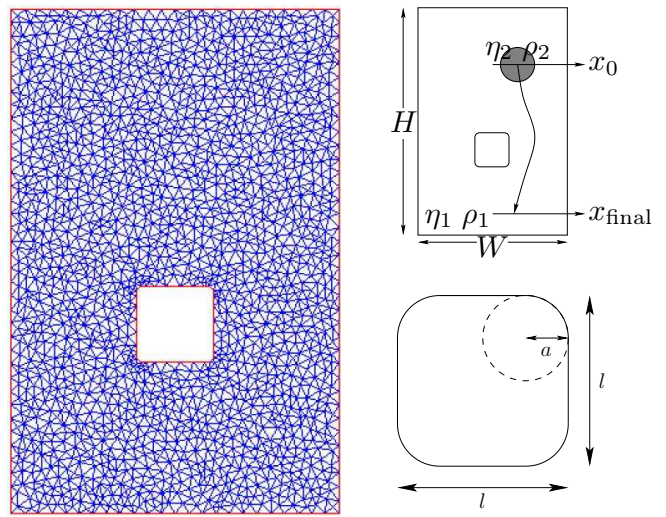}
\caption{\label{fig:geometry}For the
test study we use the geometry and mesh shown in the figure. The
general shape of the obstacle is as shown in the lower inset on
the right. The radius $a$ of the rounded corner was changed from
one simulation to the next. The aspect size of the obstacle is
$l$. The height of the channel is $H=(20/3)l$ and the width of
the channel is $W=(13/3)l$. The upper inset on the right shows
the general idea of the test study: The particles start in the
initial position $x_0$ and the final position $x_\mathrm{final}$
is recorded.}
\end{figure}
To test the implementation of the level set method in \femlab\
we have done a test study of a particle (a drop of high
viscosity and surface tension) which is passively convected in a
two dimensional fluid flow. The viscosity $\eta_2$ of the
particle was $100$ times larger than the viscosity $\eta_1$ of
the fluid. The density $\rho_1$ of the fluid was equal to the
density $\rho_2$ of the particle. The complete list of
parameters is given in Table~\ref{tab:parametervalues}.

The physical domain is an infinitely wide and infinitely long
channel with an obstacle in the center as shown in
Fig.~\ref{fig:geometry}. The boundary conditions on the fluid
are no-stress on the sides of the computational domain and
no-slip at the obstacle. The fluid velocity field is periodic
from top to bottom of the domain and is driven by a pressure
difference $\Delta p$.

We ran a series of simulations with the shape of the obstacle
changing from circular to quadratic by changing the radius of
the rounded obstacle corner $a$. Each simulation consisted of a
series of runs with different initial horizontal position $x_0$
of the particles and the initial vertical position of the
particles was $y_0=H-l$ from the top of the channel. When the
center of a convected particle is $l$ from the bottom of the
channel the final horizontal position $x_\text{final}$ is
detected (Fig.~\ref{fig:geometry}).

We represent the particle by the negative part of a level set
function and the surrounding fluid is identified by the positive
part of the level set function. The initial level set function
is given by
\begin{align}
\phi(x,y,t=0) = \sqrt{(x-x_0)^2+(y-y_0)^2}-r_p\text{,}
\end{align}
where $(x_0,y_0)$ is the initial position of the particle and
$r_p$ is the radius of the particle. Using these parameters we
solve the problem by first evolving the dynamical equations in a
small time step $\Delta t$ and then reinitialize the level set
function using the reinitialization procedures described above.
With the reinitialized level set function as initial condition
for $\phi$ we evolve the dynamical system one more time step.
This sequence is continued until the particle has moved all the
way through the system.

\section{\label{sec:results}Results}
\begin{figure}[t]
\centering
\includegraphics*[width=\columnwidth,clip]{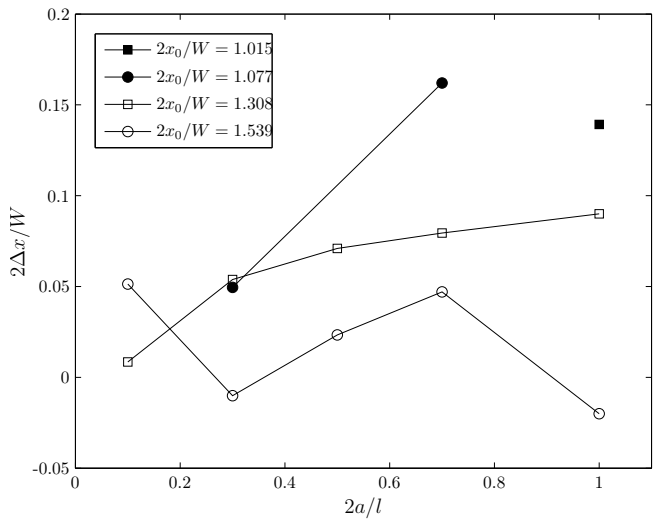}
\caption{\label{fig:deltax}%
For particles passing obstacles of different shapes normalized
difference $2\Delta x/W$ in horizontal position from start to
finish is plotted versus starting position $2a/l$. The missing
data points for the simulations with the initial positions of
the particles nearest to the center of the channel is due to the
particles getting stuck at the obstacle and hence not reaching
the final position.}
\end{figure}
We carried out simulations for four different initial positions
of the particle. The initial horizontal positions $2x_0/W$ were
$0.015$, $0.077$, $0.308$ and $0.539$, respectively. For each of
these initial positions we used five different radii of the
rounded corner of the obstacle: $2a/l = i/10,\, \text{with } i =
1,3,5,7,10$.
\begin{figure}[t]
\centering
\includegraphics*[width=\columnwidth,clip]{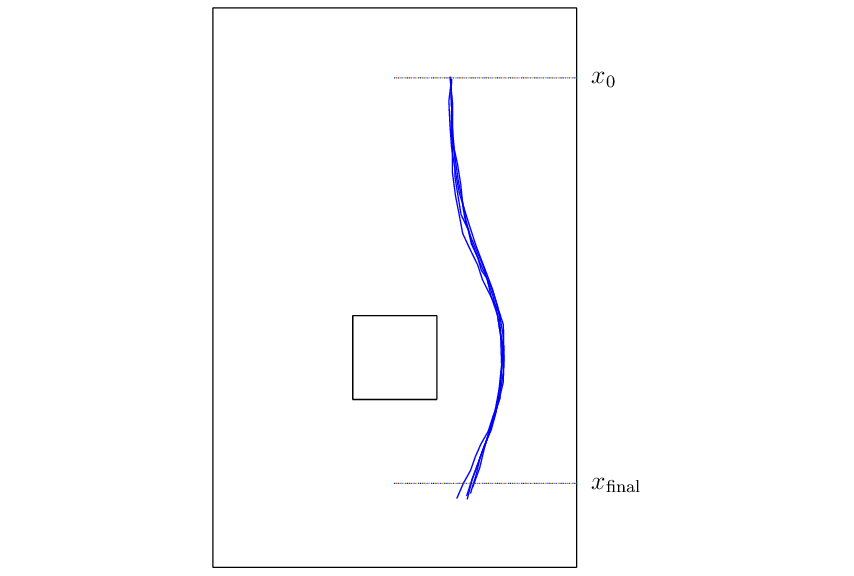}
\caption{\label{fig:path}%
The paths of particles passing obstacles of different shapes
when the starting point is $2x_0/W = 0.308$ right of the
centerline of the channel.}
\includegraphics*[width=0.45\columnwidth,clip]{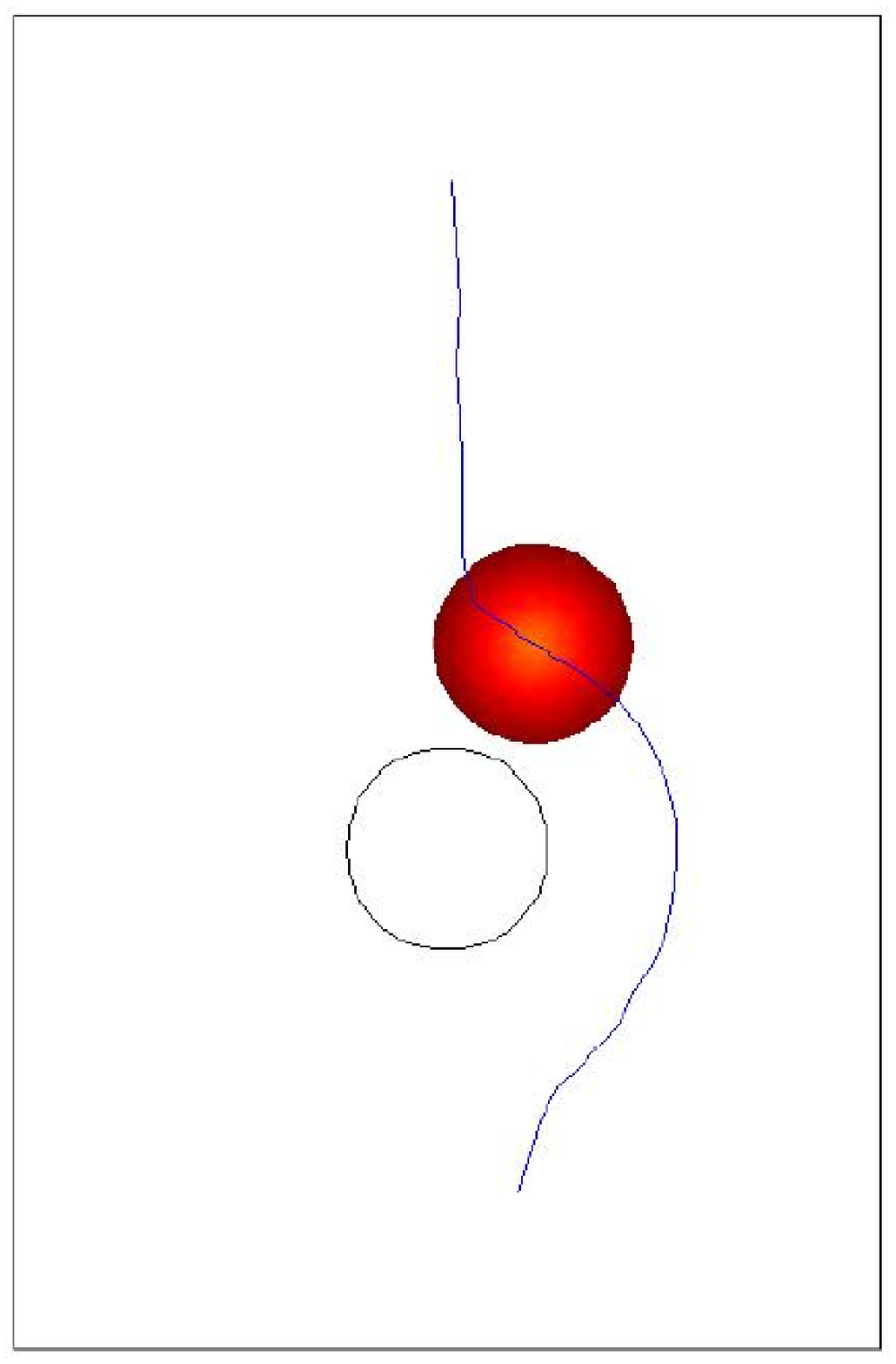}
\caption{\label{fig:path6_10}%
The path of the particle started at $2x_0/W=0.015$ when the
radius of the rounded obstacle corner is $a=l/2$. The particle
(black dot) is shown when it \lq interacts\rq\ with the
obstacle. The small gap between the particle and the obstacle
wall is caused by the smearing of the particle interface.}
\end{figure}

For each combination of initial position and obstacle shape we
solved the system and obtained the particle paths. Examples are
shown in Figs.~\ref{fig:path} and~\ref{fig:path6_10}. It is seen
that the paths of particles with the same initial position
changes as function of the shape of the obstacle
(Fig.~\ref{fig:path}). In Fig.~\ref{fig:deltax} we have plotted
the difference in the horizontal position $\Delta x$ from start
to finish.

The difference in horizontal position is almost zero for the
particles started in at the greatest distance from the center of
the channel, independent of the shape of the obstacle. As the
initial position gets closer to the center of the channel the
difference in horizontal position becomes larger and the round
obstacles tend not to drag as much in the particles as the
square obstacles yielding a larger difference in the horizontal
position.

Fig.~\ref{fig:path6_10} shows that our implementation of the
level set method is capable of coping with the interaction
forces between the stable obstacles and the moving particles
automatically.

\section{Discussion and conclusions}
\label{sec:conclusion}%
We have shown that the level set method is easily implementable in
\femlab\ and that it is a suitable method for coping with the
interaction forces between particles and hard walls automatically.
Particles can be modelled as very viscous liquid drops and the
shape preservation can be taken care of trough an appropriate
reinitialization procedure.

We have used a simple shape preserving reinitialization method.
Further work is needed in order to convect particles of an
arbitrary fixed shape. One promising reinitialisation scheme is
the particle level set method suggested by Enright
\emph{et~al}.~\cite{bib:enright}.

The level set method might prove useful when simulating
microfluidic systems for particle handling. In this paper we
have only considered the forces exerted on the particles by the
convecting fluid and thereby indirectly the forces from the
solid walls. However also other forces such as DEP forces or
magnetic forces could be taken into account making the method
applicable for simulations of many lab-on-a-chip systems
fabricated today.

\appendix
\section{} We demonstrate how to construct the
\label{app:proofs}transverse level set function $\psi$ with the
required properties. We start by defining a coordinate
transformation by
\begin{subequations}
\begin{align}
\deriv{}{\tau}\stau&=\bnabla\phi\stau\text{,}\label{eq:coordinate transformation}\\
\intertext{where}
\Big(x(s,0),y(s,0)\Big)&=\Big(\overline{x}(s),\overline{y}(s)\Big)\text{.}
\end{align}
\end{subequations}
Because of the $\delta$ function in Eq.~\eqref{eq:the level set
equation} $\psi$ only needs to fulfill the requirements in a small
region $|\tau|<\epsilon$ around $\Gamma$. In this small region we
can define $\psi$ as
\begin{equation}
\psi\stau = \psi_0(s)\text{,}
\end{equation}
where $\psi_0(s)$ is a smooth increasing function if and only if the mapping
of $(x,y)$ to $(s,\tau)$ is one-to-one. Using the change of
variables theorem we have to show that
\begin{align}\label{eq:determinant}
\left|\pderiv{(x,y)}{(s,\tau)}\right|\neq 0\text{.}
\end{align}
Taylor expanding Eq.~\eqref{eq:coordinate transformation}
around $\tau=0$ yields
\begin{align}
(x_\tau,y_\tau)=\bnabla\phi\Big(\overline{x}(s),\overline{y}(s)\Big)+\order{\tau}\text{.}
\end{align}
Differentiation of Eq.~\eqref{eq:coordinate transformation}
with respect to $s$ and integration with respect to $\tau$ yields
\begin{multline}
\int_{0}^{\tau}\!\deriv{}{s}\deriv{}{\tau'}\stau[']\diff \tau'=\\
\int_{0}^{\tau}\!\deriv{}{s}\bnabla\phi\stau[']\diff \tau'
\end{multline}
From which follows
\begin{multline}
\Big(x_s(s,\tau),y(s,\tau)\Big)-\Big(x_s(s,0),y_s(s,0)\Big)=\\
\int_{0}^{\tau}\!\deriv{}{s}\bnabla\phi\stau[']\diff\tau'\text{,}
\end{multline}
and thus
\begin{align}
\Big(&x_s(s,\tau),y(s,\tau)\Big)\nonumber\\
&=\Big(\overline{x}_s(s),\overline{y}_s(s)\Big)+\int_{0}^{\tau}\!\deriv{}{s}\bnabla\phi\stau[']\diff\tau'\nonumber\\
&=\mathbf{T}(s)+\order{\tau}\text{.}
\end{align}
Here $\mathbf{T}$ is a unit tangent vector to the interface. We
can now calculate the determinant~\eqref{eq:determinant}
\begin{align}
\begin{split}
\determinant{\pderiv{(x,y)}{(s,\tau)}}
&=(x_\tau,y_\tau)\cdot(-y_s,x_\tau)\\
&=\bnabla\phi(\overline{x}_s,\overline{y}_s)\cdot\hat{\mathbf{T}}\\
&=|\bnabla\phi| |\mathbf{T}|+\order{\tau}\\
&=|\bnabla\phi|_{\phi=0}+\order{\tau}\neq 0\text{.}
\end{split}
\end{align}
This means that $\psi$ is well defined in a small region around
$\Gamma$. Now all we need to prove is that $\bnabla\phi$ and
$\bnabla\psi$ are orthogonal and that $|\bnabla\psi|\neq 0$. The
orthogonality can be proved by differentiating $\psi$ with respect
to $\tau$,
\begin{align}
\begin{split}
\deriv{}{\tau}\psi\stau &=\psi_x x_\tau + \psi_y y_\tau\\
&=\bnabla\psi\cdot\bnabla\phi=\deriv{\psi_0(s)}{\tau}=0\text{,}
\end{split}
\end{align}
which means that $\phi$ and $\psi$ are orthogonal if and only if
$|\bnabla\psi|\neq 0$. This follows immediately from
differentiating $\psi$ with respect to $s$,
\begin{align}
\begin{split}
\deriv{}{s}\psi\stau &=\psi_x x_s+\psi_y y_s\\
&=\bnabla\psi\cdot(x_s,y_s)\\
&=\bnabla\psi\cdot\mathbf{T}\\
&=|\bnabla\psi|=\psi'_0(s)>0\text{,}
\end{split}
\end{align}
because $\psi_0(s)$ was chosen to be an increasing function.
Thereby we have established the level set formulation of the
Navier--Stokes equation for a two liquid flow of incompressible
fluids.


\begin{thebibliography}{10}
\expandafter\ifx\csname
natexlab\endcsname\relax\def\natexlab#1{#1}\fi
\expandafter\ifx\csname bibnamefont\endcsname\relax
  \def\bibnamefont#1{#1}\fi
\expandafter\ifx\csname bibfnamefont\endcsname\relax
  \def\bibfnamefont#1{#1}\fi
\expandafter\ifx\csname citenamefont\endcsname\relax
  \def\citenamefont#1{#1}\fi
\expandafter\ifx\csname url\endcsname\relax
  \def\url#1{\texttt{#1}}\fi
\expandafter\ifx\csname
urlprefix\endcsname\relax\def\urlprefix{URL }\fi
\providecommand{\bibinfo}[2]{#2}
\providecommand{\eprint}[2][]{\url{#2}}

\bibitem[{\citenamefont{Chou et~al.}(1999)\citenamefont{Chou, Bakajin, Turner,
  Duke, Chan, Cox, Craighead, and Austin}}]{bib:chou}
\bibinfo{author}{\bibfnamefont{C.-F.} \bibnamefont{Chou}},
  \bibinfo{author}{\bibfnamefont{O.}~\bibnamefont{Bakajin}},
  \bibinfo{author}{\bibfnamefont{S.~W.~P.} \bibnamefont{Turner}},
  \bibinfo{author}{\bibfnamefont{T.~A.~J.} \bibnamefont{Duke}},
  \bibinfo{author}{\bibfnamefont{S.~S.} \bibnamefont{Chan}},
  \bibinfo{author}{\bibfnamefont{E.~C.} \bibnamefont{Cox}},
  \bibinfo{author}{\bibfnamefont{H.~G.} \bibnamefont{Craighead}},
  \bibnamefont{and} \bibinfo{author}{\bibfnamefont{R.~H.}
  \bibnamefont{Austin}}, \bibinfo{journal}{USA} \textbf{\bibinfo{volume}{96}},
  \bibinfo{pages}{13762} (\bibinfo{year}{1999}).

\bibitem[{\citenamefont{Duke and Austin}(1998)}]{bib:duke}
\bibinfo{author}{\bibfnamefont{T.~A.~J.} \bibnamefont{Duke}} \bibnamefont{and}
  \bibinfo{author}{\bibfnamefont{R.~H.} \bibnamefont{Austin}},
  \bibinfo{journal}{Phys. Rev. Lett.} \textbf{\bibinfo{volume}{80}},
  \bibinfo{pages}{1552} (\bibinfo{year}{1998}).

\bibitem[{\citenamefont{Huang et~al.}(2004)\citenamefont{Huang, Cox, Austin,
  and Sturm}}]{bib:huang04}
\bibinfo{author}{\bibfnamefont{L.~R.} \bibnamefont{Huang}},
  \bibinfo{author}{\bibfnamefont{E.~C.} \bibnamefont{Cox}},
  \bibinfo{author}{\bibfnamefont{R.~H.} \bibnamefont{Austin}},
  \bibnamefont{and} \bibinfo{author}{\bibfnamefont{J.~C.} \bibnamefont{Sturm}},
  \bibinfo{journal}{Science} \textbf{\bibinfo{volume}{304}},
  \bibinfo{pages}{987} (\bibinfo{year}{2004}).

\bibitem[{\citenamefont{Huang et~al.}(2002)\citenamefont{Huang, Silberzan,
  Tegenfeldt, Cox, Sturm, Austin, and Craighead}}]{bib:huang02}
\bibinfo{author}{\bibfnamefont{L.~R.} \bibnamefont{Huang}},
  \bibinfo{author}{\bibfnamefont{P.}~\bibnamefont{Silberzan}},
  \bibinfo{author}{\bibfnamefont{J.}~\bibnamefont{Tegenfeldt}},
  \bibinfo{author}{\bibfnamefont{E.~C.} \bibnamefont{Cox}},
  \bibinfo{author}{\bibfnamefont{J.~C.} \bibnamefont{Sturm}},
  \bibinfo{author}{\bibfnamefont{R.~H.} \bibnamefont{Austin}},
  \bibnamefont{and}
  \bibinfo{author}{\bibfnamefont{H.}~\bibnamefont{Craighead}},
  \bibinfo{journal}{Phys. Rev. Lett.} \textbf{\bibinfo{volume}{89}},
  \bibinfo{pages}{1} (\bibinfo{year}{2002}).

\bibitem[{\citenamefont{Sethian}(1999)}]{bib:sethian}
\bibinfo{author}{\bibfnamefont{J.~A.} \bibnamefont{Sethian}},
  \emph{\bibinfo{title}{Level Set Methods and Fast Marching Methods}}
  (\bibinfo{publisher}{Cambridge University Press}, \bibinfo{year}{1999}),
  \bibinfo{edition}{2nd} ed.

\bibitem[{\citenamefont{Chang et~al.}(1996)\citenamefont{Chang, Hou, Merriman,
  and Osher}}]{bib:chang}
\bibinfo{author}{\bibfnamefont{Y.~C.} \bibnamefont{Chang}},
  \bibinfo{author}{\bibfnamefont{T.~Y.} \bibnamefont{Hou}},
  \bibinfo{author}{\bibfnamefont{B.}~\bibnamefont{Merriman}}, \bibnamefont{and}
  \bibinfo{author}{\bibfnamefont{S.}~\bibnamefont{Osher}},
  \bibinfo{journal}{J.~Comput. Phys.} \textbf{\bibinfo{volume}{124}},
  \bibinfo{pages}{449} (\bibinfo{year}{1996}).

\bibitem[{\citenamefont{\femlab\ homepage}()}]{bib:comsol}
\bibinfo{author}{\bibnamefont{\femlab\ homepage}},
  \bibinfo{note}{\texttt{www.comsol.dk}}.

\bibitem[{\citenamefont{Osher and Fedkiw}(2003)}]{bib:osher}
\bibinfo{author}{\bibfnamefont{S.}~\bibnamefont{Osher}} \bibnamefont{and}
  \bibinfo{author}{\bibfnamefont{R.}~\bibnamefont{Fedkiw}},
  \emph{\bibinfo{title}{Level Set Methods and Dynamic Implicit Surfaces}}, vol.
  \bibinfo{volume}{153} of \emph{\bibinfo{series}{Applied mathematical
  sciences}} (\bibinfo{address}{Springer-Verlag New York},
  \bibinfo{year}{2003}), \bibinfo{edition}{1st} ed.

\bibitem[{\citenamefont{Sussman and Fatemi}(1999)}]{bib:sussmann}
\bibinfo{author}{\bibfnamefont{M.}~\bibnamefont{Sussman}} \bibnamefont{and}
  \bibinfo{author}{\bibfnamefont{E.}~\bibnamefont{Fatemi}},
  \bibinfo{journal}{SIAM J.~Sci. Comput.} \textbf{\bibinfo{volume}{20}},
  \bibinfo{pages}{1165} (\bibinfo{year}{1999}).

\bibitem[{\citenamefont{Enright et~al.}(2002)\citenamefont{Enright, Fedkiw,
  Ferziger, and Mitchell}}]{bib:enright}
\bibinfo{author}{\bibfnamefont{D.}~\bibnamefont{Enright}},
  \bibinfo{author}{\bibfnamefont{R.}~\bibnamefont{Fedkiw}},
  \bibinfo{author}{\bibfnamefont{J.}~\bibnamefont{Ferziger}}, \bibnamefont{and}
  \bibinfo{author}{\bibfnamefont{I.}~\bibnamefont{Mitchell}}
  (\bibinfo{year}{2002}).

\end{thebibliography}
\end{document}